\documentclass[a4paper]{jpconf}
\usepackage{graphicx}
\begin{document}
\title{Strangeness in Quark Matter: Opening Talk}

\author{J.~Steinheimer$^1$, T.~Lang$^1$, H.~van Hees$^1$, A.S.~Botvina$^{2,3}$, K.K.~Gudima$^4$, I.N.~Mishustin$^{1,5}$, H.~St\"ocker$^{1,6,7}$ and M.~Bleicher$^{1,7}$}

\ead{steinheimer@fias.uni-frankfurt.de}

\address{$^1$ Frankfurt Institute for Advanced Studies,
Johann Wolfgang Goethe\\Universit\"at,
Ruth-Moufang-Str. 1, 60438 Frankfurt am Main, Germany\\
$^2$Institute for Nuclear 
Research, Russian Academy of Sciences, 117312 Moscow, Russia
$^3$Helmholtz-Institut Mainz, J.Gutenberg-Universit\"at, 55099 Mainz, Germany\\
$^4$Institute of Applied Physics, Academy of Sciences of Moldova, 
MD-2028 Kishinev, Moldova\\
$^5$Kurchatov Institute, Russian Research Center,
123182 Moscow, Russia\\
$^6$GSI Helmholtzzentrum f\"ur Schwerionenforschung GmbH, Planckstr. 1, 
64291 Darmstadt, Germany\\
$^7$Institut f\"ur Theoretische Physik, Goethe Universit\"at, Max-von-Laue-Str. 1,
60438 Frankfurt am Main, Germany\\
}
\begin{abstract}
We discuss several new developments in the field of strange and heavy flavor physics in high energy heavy ion collisions. As shown by many recent theoretical works, heavy flavored particles give us a unique opportunity to study the properties of systems created in these collisions. 
Two in particular important aspects, the production of (multi) strange hypernuclei and the properties of heavy flavor mesons, are at the core of several future facilities and will be discussed in detail.

\end{abstract}

As strange quarks have to be newly produced during the hot and dense stage of a relativistic nuclear collision, 
they carry information, on the properties of the matter that was 
created \cite{Koch:1986ud,rafelski82,rafmull82,birozim}, to the observed final particle state. The enhancement of strange particle production is discussed \cite{Adams:2005dq,Back:2004je,Arsene:2004fa,Gazdzicki:2004ef,Gazdzicki:1998vd} as a possible signal for the creation of a deconfined phase.
Recently several observables, regarding strange and charm quarks have shown the importance of understanding the dynamics of strangeness and charm production in heavy ion collisions:

\begin{itemize}
\item Strange particle ratios and yields from the ALICE collaboration may indicate that there is either no unique chemical freeze out point for strange an non-strange particles \cite{Bellwied:2013cta,Becattini:2012xb,Milano:2013sza,Bass:2000ib}, or the light quark phase space is severely over-saturated \cite{Petran:2013lja}.
\item Lattice calculations on the stability of the H-dibaryon indicate it might be either very loosely bound or a resonant state \cite{Beane:2011zpa,Inoue:2010es,Buchoff:2012ja}.
\item Viscous hydrodynamics, with fluctuating initial conditions \cite{Bleicher:1998wd} and finite but small viscous corrections, seems to describe strange hadron observables even at large baryon densities \cite{Karpenko:2013ksa,Petersen:2006vm}. 
\item The hydrodynamic model calculations show sensitivity on the life time of the system and the applied equation of state \cite{Li:2008qm}. 
\item There are indications that systems created in high energy p+p and p+Pb collisions can thermalize/equilibrate to a certain degree and show signs of collectivity \cite{Vogel:2010et,Lang:2013ex,Werner:2013ipa,Bozek:2013ska}.
\item A polarization of $\Lambda$'s due to the finite angular momentum of the fireball is expected \cite{Becattini:2013vja}.
\item Possible signals for the observation of quarkionic matter where proposed \cite{Torrieri:2013mk}.
\item There still is the unexplained puzzle of the strongly enhanced $\Xi^-$ yield at the HADES experiment \cite{Agakishiev:2010rs}.
 \end{itemize}

In addition to these interesting results we will focus on two topics which are at the core of present and upcoming experimental programs, FAIR with CBM \cite{Senger:2011zza} and NICA, namely the production of strange clusters and hypernuclei and the description of heavy quark transport in nuclear collisions.

\section{Hypernuclei}
Although abundantly produced, the interactions of strange hadrons are not very
well understood but important for the description of 
the hadronic phase of a heavy ion collision and dense hadronic matter as can be found in compact stars. 
Hypernuclear physics offers a direct
experimental way to study hyperon--nucleon ($YN$) and hyperon--hyperon
($YY$) interactions.\\
Exotic forms of deeply bound objects with strangeness \cite{Bodmer:1971we} and later the H di-baryon \cite{Jaffe:1976yi} was proposed by theory. Recent lattice QCD calculations suggest that the 
H-dibaryon is either a weakly bound system or a resonant state \cite{Beane:2010hg,Beane:2011xf,Inoue:2010es,Buchoff:2012ja}, and there could be strange di-baryon systems including $\Xi$'s that can be bound \cite{Beane:2011iw}. An experimental confirmation of such a state would therefore be an enormous advance in the understanding of the hyperon interaction.\\
Hypernuclei are known to be produced in heavy ion collisions already 
for a long time \cite{nucl-th/9412035,Ahn:2001sx,Takahashi:2001nm,arXiv:1010.2995}, 
and the recent discoveries of the first anti-hypertriton \cite{star2010} and anti-$\alpha$ \cite{star2011} has fueled the interest in the field of hypernuclear physics.
One can discriminate two 
distinct mechanisms for hypercluster formation in heavy ion collisions. First, 
the absorption of hyperons in the spectator fragments of non central  
collisions \cite{Ko:1985gp,Gaitanos:2007mm,Gaitanos:2009at,Botvina:2011jt}. 
The hyper-systems obtained here are rather large and moderately excited, decaying into hyper fragments later on \cite{Botvina:2011jt,Botvina:2007pd}.

\begin{figure}[t]
\begin{minipage}[t]{0.47\textwidth}
\includegraphics[width=\textwidth]{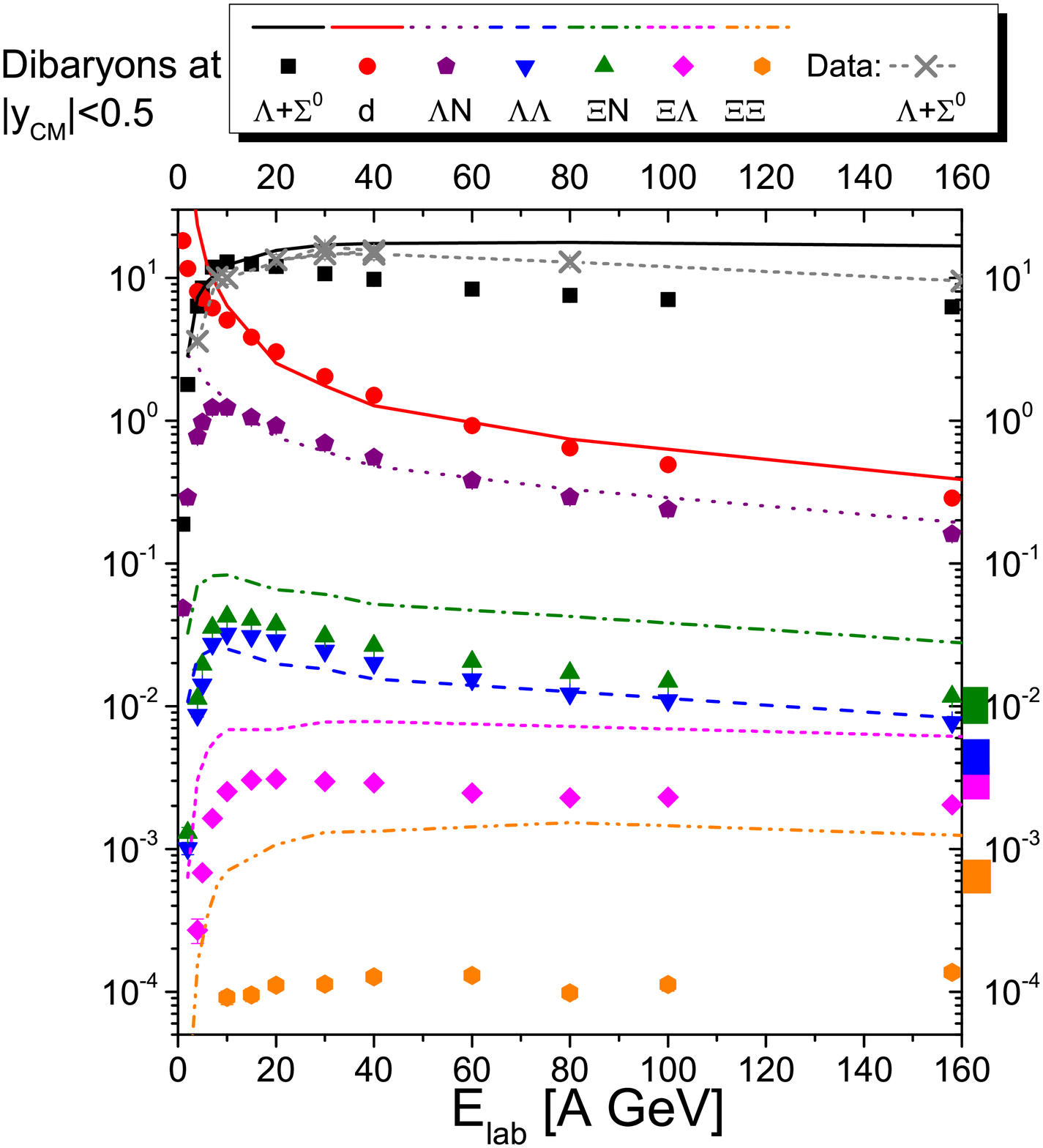}
\caption{Yields of di-baryons in the mid rapidity region ($|y|<0.5$) of most central collisions of Pb+Pb/Au+Au. Thermal production in the UrQMD hybrid model (lines) is compared to coalescence results with the DCM model (symbols). Black lines and symbols depict the production of $\Lambda$'s from both models, compared to data (grey crosses) from \cite{Ahmad:1991nv,Mischke:2002wt,Alt:2008qm}.
\label{dibmidy}}
\end{minipage}\hspace{0.03\textwidth}
\begin{minipage}[t]{0.47\textwidth}
\includegraphics[width=\textwidth]{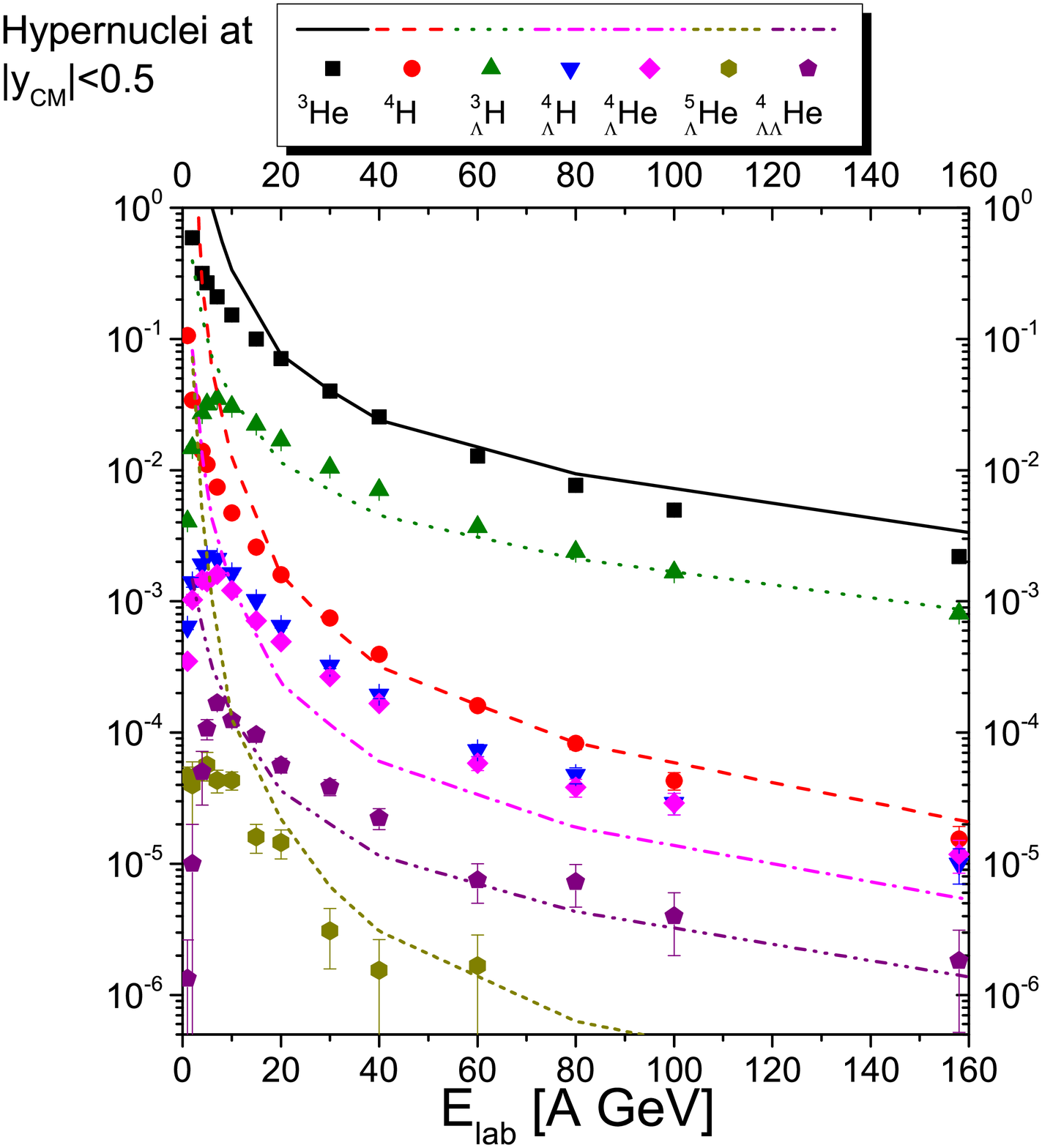}
\caption{Yields per event of different (hyper-)nuclei in the mid rapidity region ($|y|<0.5$) of most central collisions of Pb+Pb/Au+Au. Shown are the results from the thermal production in the UrQMD hybrid model (lines) as compared to coalescence results with the DCM model (symbols). 
\label{hypmidy}
}
\end{minipage}\hspace{0.03\textwidth}%
\end{figure}

Alternatively, (hyper-)nuclear clusters can emerge from the hot and dense fireball 
region of the reaction. In this scenario the cluster is formed at, or shortly after, 
the (chemical-)freeze out of the system. A general assumption is, that these 
clusters are then formed through coalescence of different newly produced 
hadrons \cite{Scheibl:1998tk}. To estimate the production yield we compare two distinct approaches.
First we use the hadronic transport model DCM \cite{Toneev:1983cb} to provide us with the phase space information of 
all hadrons produced in a heavy ion collision. This information then serves as an 
input for a coalescence prescription \cite{Steinheimer:2012tb}. On the other hand it has been shown that thermal models consistently describe the production yields of hadrons (and nuclei 
\cite{Andronic:2008gu}) very well. We can therefore assume thermal production of clusters from a fluid 
dynamical description to heavy ion collisions \cite{Steinheimer:2008hr,Petersen:2008dd}.
Figures \ref{dibmidy} and \ref{hypmidy} show the results for di-baryon and hypernuclei yields in the mid-rapidity region of central, $b<3.4$ fm, heavy ion collisions at different beam energies. We compare results from the coalescence approach (symbols) with those from the fluid dynamical model (solid lines). It is clearly visible that large and multi-strange nuclear clusters exhibit a production maximum at a lower beam energy of $E_{lab} =10 A$ GeV. This means that experiments at the future facilities in Dubna and at FAIR will be well suited for the search of these new and exotic clusters.

\begin{figure}[t]
\begin{minipage}[t]{0.44\textwidth}
\includegraphics[width=\textwidth]{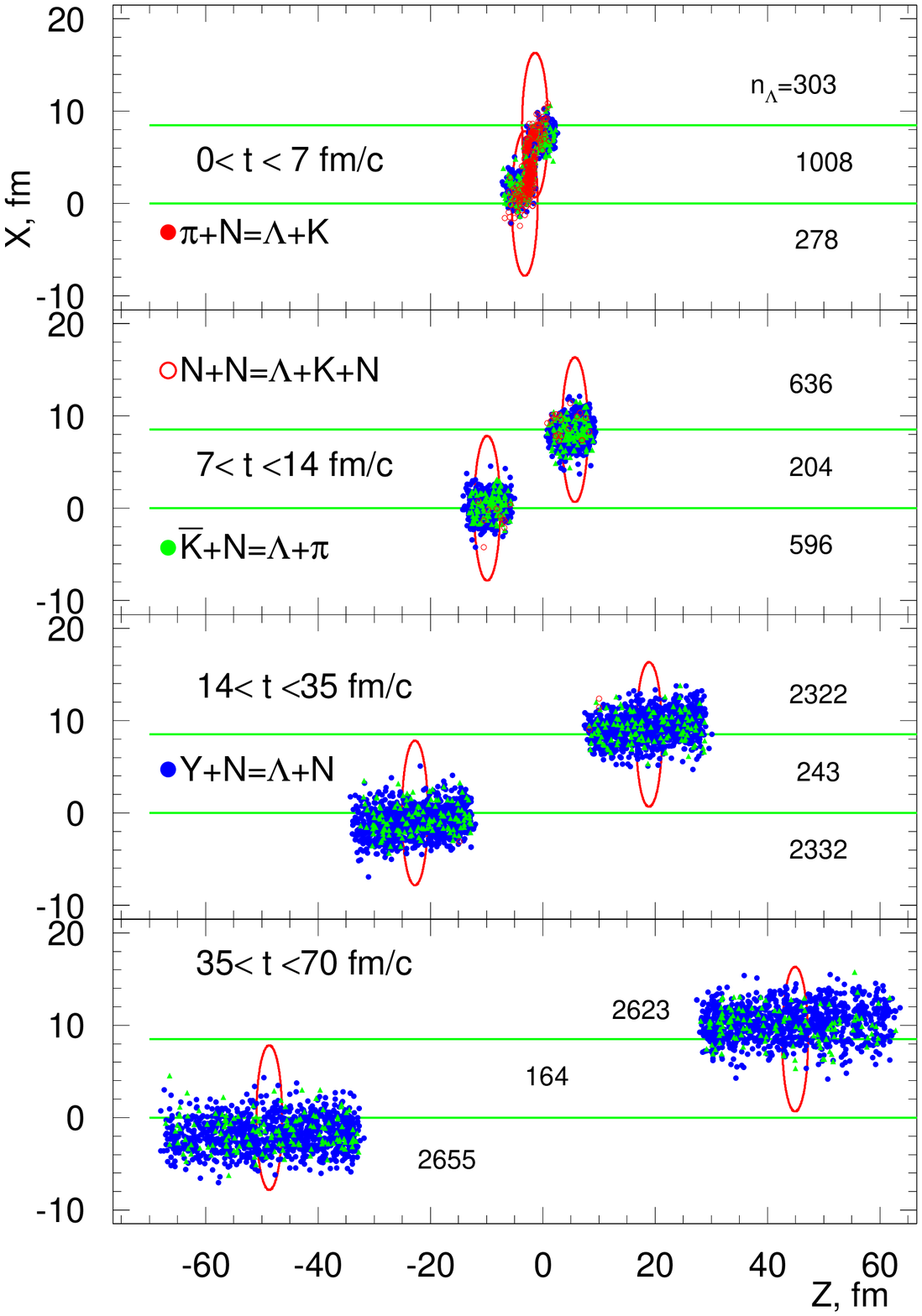}
\caption{Coordinates of $\Lambda$ absorption in the X--Z plane.
The number of hyperons $n_{\Lambda}$ (per 2$\cdot$10$^5$ events) captured in the 
participant and spectator zones during these intervals is noted on the 
right side.}\label{fig3}
\end{minipage}\hspace{0.03\textwidth}
\begin{minipage}[t]{0.50\textwidth}
\includegraphics[width=\textwidth]{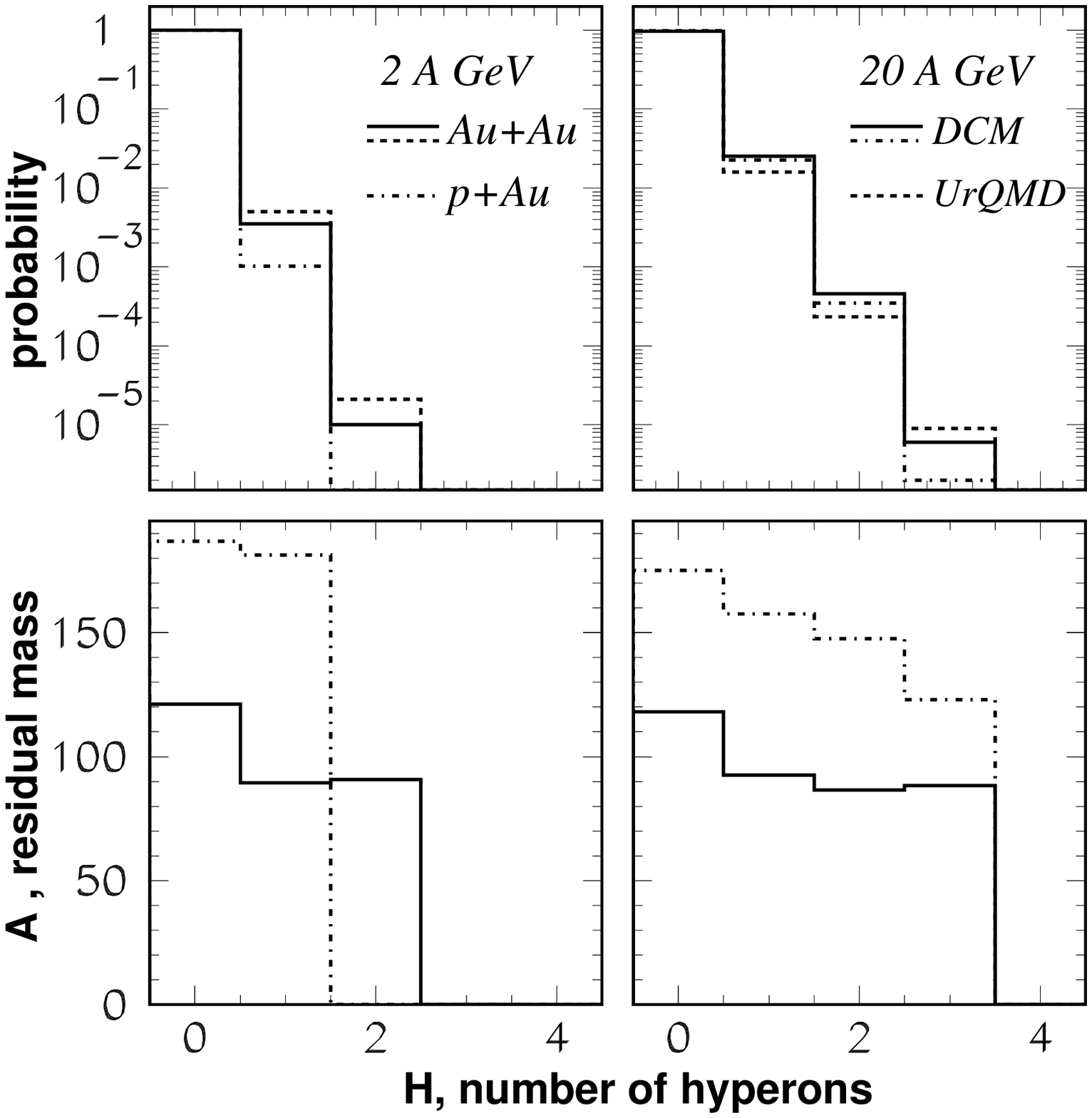}
\caption{Formation probability of strange spectator residuals 
(top), and their mean mass numbers (bottom)
versus the number of captured $\Lambda$ hyperons (H), calculated with DCM 
and UrQMD model for p + Au and  Au + Au collisions with energy of 
2 GeV (left), and 20 GeV per nucleon (right). }\label{fig6}
\end{minipage}\hspace{0.03\textwidth}%
\end{figure}
 
$\Lambda$-hyperons are produced mainly in the hot and dense fireball, however, they 
have a broad rapidity distribution, so that a certain fraction of them can
 be found in the spectator kinematical region. Some of these $\Lambda$-hyperons
 are captured by nuclear spectator 
fragments produced in peripheral collisions \cite{Gaitanos:2007mm}. 
The production of large excited spectator residues is well established in 
relativistic heavy-ion collisions\cite{Botvina:1994vj}. 
We expect that the capture of hyperons by spectators leads to 
their exitation and break-up into fragments \cite{Botvina:2007pd}.
Using the DCM \cite{Toneev:1983cb} and UrQMD \cite{Bleicher:1999xi,Bass:1998ca} models we simulate peripheral (b=$8.5$ fm) relativistic nuclear reactions to calculate the local nucleon density $\rho$ at the positions of the hyperons created.
This local density is then used to determine the 
effective potential $V_{\Lambda} (\rho )=-\alpha \frac{\rho}{\rho_0}\left[1-\beta (\frac{\rho}{\rho_0})^{2/3}\right]$, parameterized in Ref.~\cite{Ahmad:1983re}. Here $\alpha=57.5$ MeV, and $\beta =0.522$. A $\Lambda$ can be considered
as absorbed when its kinetic energy, relative to the spectator, is smaller then the binding energy due to the nuclear density. 
Figure \ref{fig3} shows the space coordinates of the absorption of $\Lambda$'s, by the spectators, for different time intervals, calculated with the DCM transport model. One can clearly see how the $\Lambda$'s are absorbed in the spectator region.
In figure \ref{fig6} we show the probabilities for the formation of a strange spectator residual for two different beam energies, calculated with the DCM and UrQMD transport model. Both models predict residuals with even 3 absorbed $\Lambda$'s.

\section{Charm Transport}

\begin{figure}[t]
\begin{minipage}[t]{0.47\textwidth}
\includegraphics[width=\textwidth]{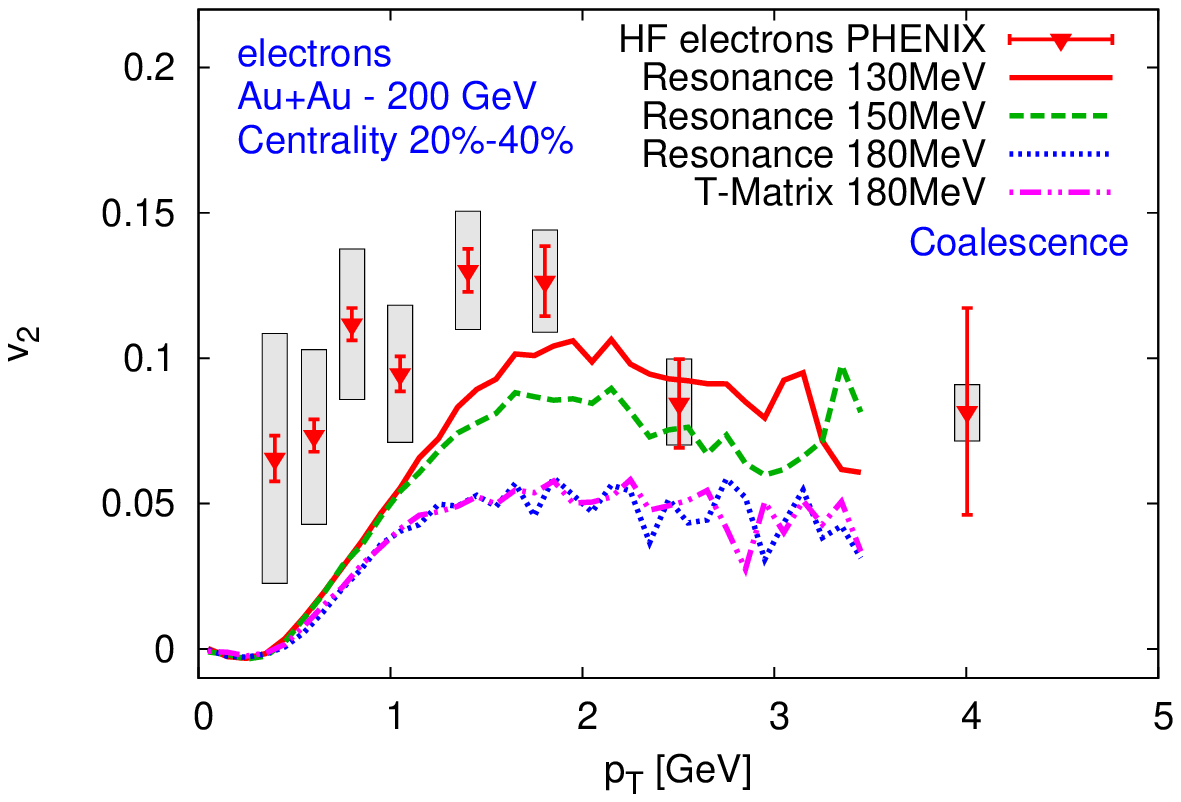}
\caption{Elliptic flow $v_2$ of electrons from heavy quark decays
  in Au+Au collisions at $\sqrt{s_{NN}}=200$ GeV using a coalescence
  mechanism, compared to data \cite{Adare:2010de}. We use a rapidity cut of $|y|<0.35$.}\label{flowRHIC4}
\end{minipage}\hspace{0.03\textwidth}
\begin{minipage}[t]{0.47\textwidth}
\includegraphics[width=\textwidth]{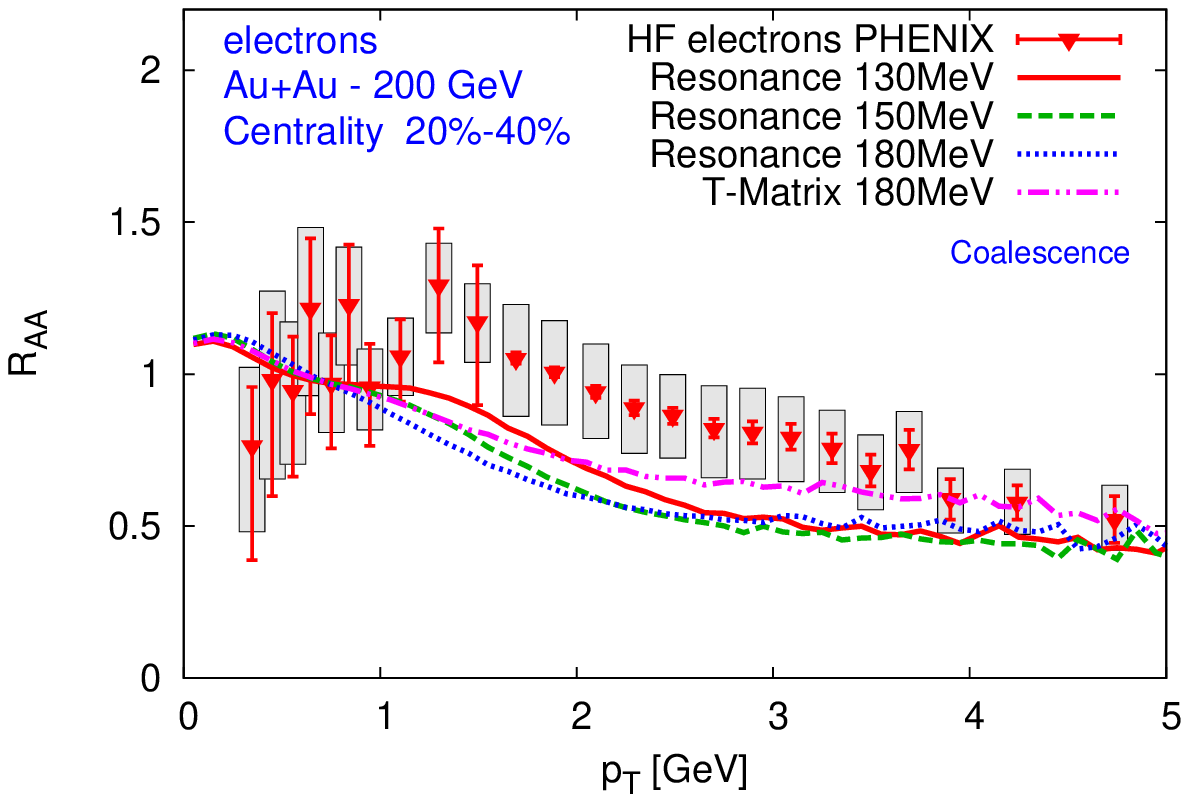}
\caption{Nuclear modification factor 
$R_{AA}$ of electrons from heavy quark decays
  in Au+Au collisions at $\sqrt {s_{NN}}=200$ GeV using a coalescence
  mechanism, compared to data \cite{Adare:2010de}. We use a rapidity cut of $|y|<0.35$.}\label{RAARHIC4}
\end{minipage}\hspace{0.03\textwidth}%
\end{figure}

\begin{figure}[t]
\begin{minipage}[t]{0.47\textwidth}
\includegraphics[width=\textwidth]{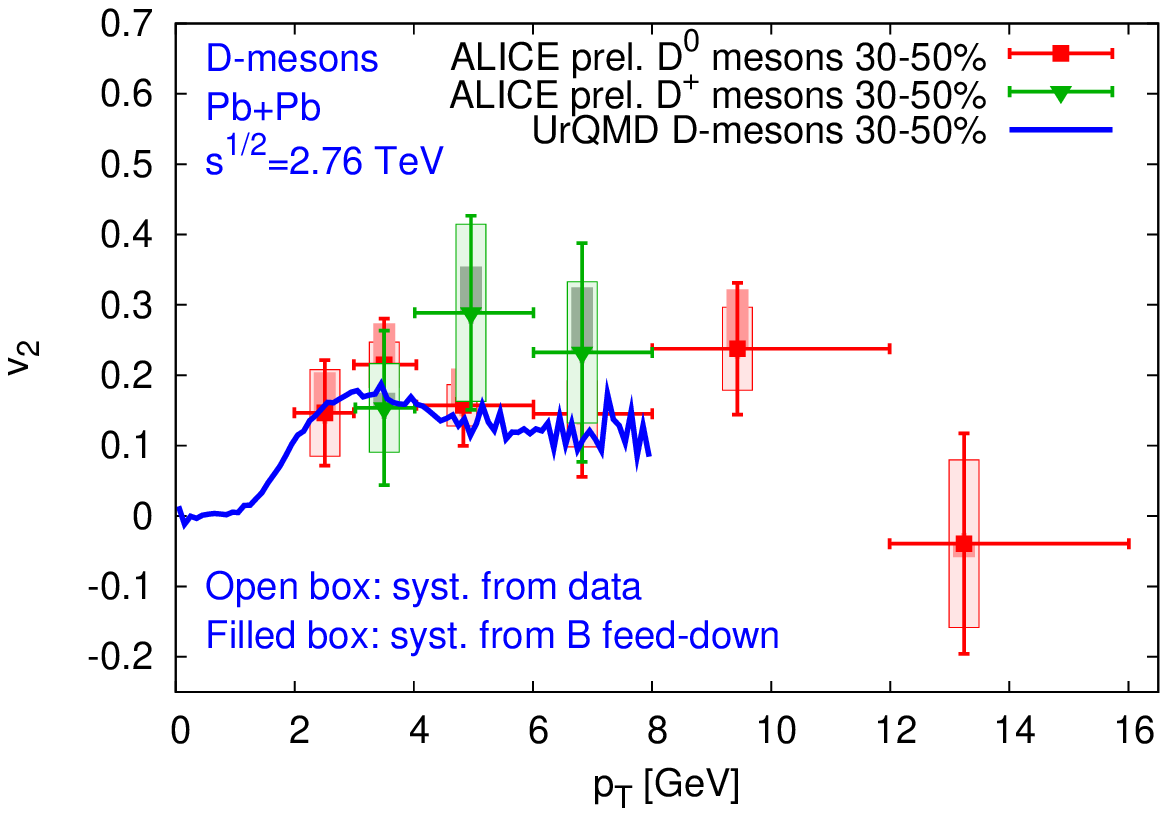}
\caption{Flow $v_2$ of D-mesons in Pb+Pb collisions at $\sqrt
  {s_{NN}}=2.76$ TeV compared to data from the ALICE experiment \cite{Abelev:2013lca}. 
A rapidity cut of $|y|<0.35$ is employed.}\label{flowLHC4}
\end{minipage}\hspace{0.03\textwidth}
\begin{minipage}[t]{0.47\textwidth}
\includegraphics[width=\textwidth]{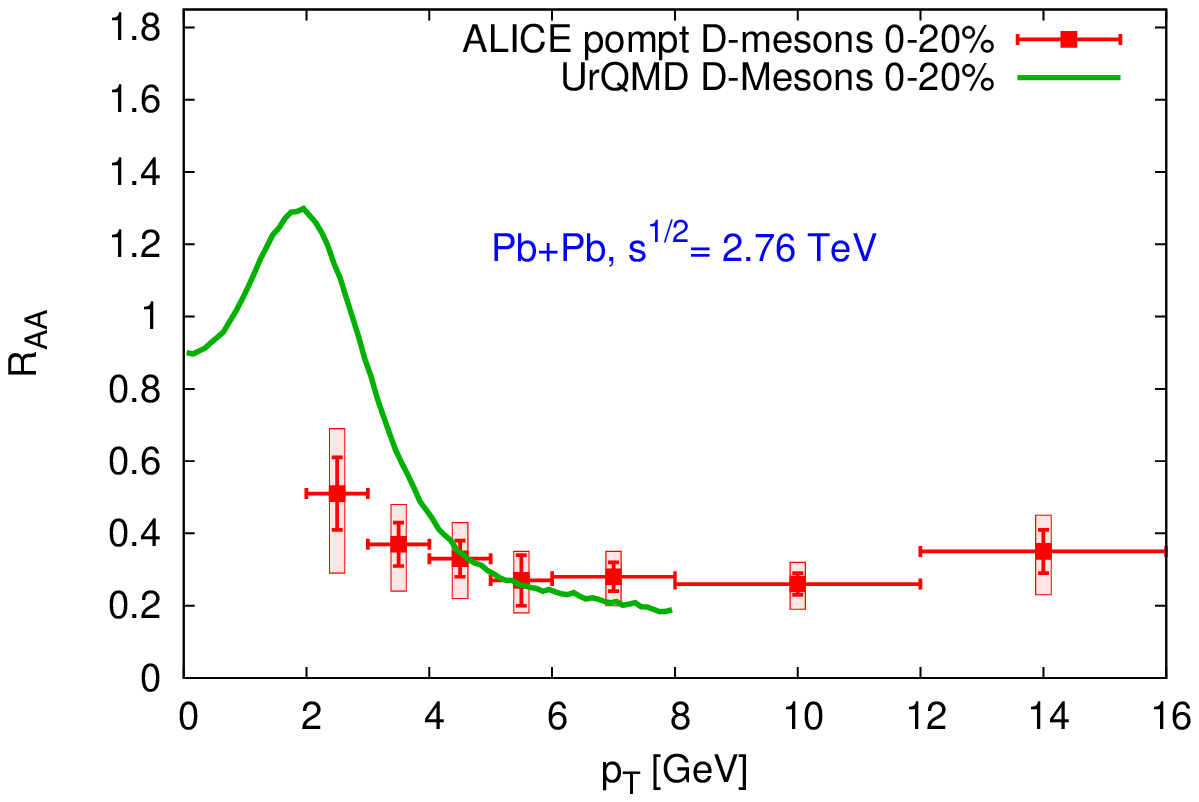}
\caption{$R_{AA}$ of D-mesons in Pb~Pb collisions at $\sqrt
  {s_{NN}}=2.76$ TeV compared to experimental data from ALICE
  \cite{ALICE:2012ab}. A rapidity cut of $|y|<0.35$ is employed.}\label{RAALHC4}
\end{minipage}\hspace{0.03\textwidth}%
\end{figure}

Heavy quarks are an ideal probe for the QGP, mainly produced in the
initial hard processes of a heavy ion collision. The most
interesting observables are the nuclear modification factor, $R_{AA}$,
and the elliptic flow, $v_2$. The measured large elliptic flow, $v_2$, of
open heavy-flavor mesons indicate a high degree of thermalization of
the heavy quarks with the bulk medium. 
A quantitative analysis of the degree of thermalization of
heavy-quark degrees of freedom may lead to an understanding 
of the transport properties of QCD.
We use a hybrid model, consisting of the UrQMD model
\cite{Bass:1998ca,Bleicher:1999xi} and a full (3+1)-dimensional ideal
hydrodynamical model \cite{Rischke:1995ir,Rischke:1995mt} to simulate
the bulk medium created in an ultra relativistic nuclear collision (for alternative approaches see \cite{Uphoff:2010sh,Gossiaux:2010yx,Linnyk:2012pu}).
The heavy-quark propagation in the medium is described
by a relativistic Langevin approach \cite{Rapp:2009my}. 
Within this
framework we use different drag and
diffusion coefficients of heavy
flavors on the heavy-quark observables and compare the results with the
experimental data from the Relativistic Heavy Ion Collider (RHIC) 
and the Large Hadron Collider (LHC).
To make comparisons to experiment we coalesce a light quark with a heavy quark \cite{Lang:2012cx}.
Figures \ref{flowRHIC4} through \ref{RAALHC4} show our results on the $v_2$ and $R_{AA}$ of electrons from heavy quark decays for two different beam energies. One can see that our model is in rather good agreement with the data when a late decoupling criterion is chosen.

\section{Summary}
We have shown two important aspects of present and future experimental programs dedicated on the study of strange and heavy flavor physics. The study of hypernuclei and multi strange clusters can deepen our understanding of hyperon interactions as well as hadron formation in heavy ion collisions. Charm quarks on the other hand can serve as an important probe for the hot and dense QGP phase of the collision and are now used to extract information on the transport properties of the QCD medium.

\section{Acknowledgements}
This work has been supported by GSI and Hessian initiative for excellence (LOEWE) 
through the Helmholtz International Center for FAIR (HIC for FAIR). T.\ Lang gratefully acknowledges support from the Helmholtz Research School on Quark Matter Studies. I.M. acknowledges partial support from grant NS-215.2012.2 (Russia). The computational resources were provided by the LOEWE Frankfurt Center for Scientific Computing (LOEWE-CSC). 

\end{document}